\documentclass[prd,aps,floats,tabularx,preprintnumbers,twocolumn]{revtex4}
\usepackage{graphicx, epsfig}
\usepackage{amsmath}
\newcommand{\be}{\begin{equation}}
\newcommand{\ee}{\end{equation}}
\newcommand{\bea}{\begin{eqnarray}}
\newcommand{\eea}{\end{eqnarray}}

\def\fun#1#2{\lower3.6pt\vbox{\baselineskip0pt\lineskip.9pt
        \ialign{$\mathsurround=0pt#1\hfill##\hfil$\crcr#2\crcr\sim\crcr}}}

\newcommand\lsim{\mathrel{\rlap{\lower4pt\hbox{\hskip1pt$\sim$}}
    \raise1pt\hbox{$<$}}}
\newcommand\gsim{\mathrel{\rlap{\lower4pt\hbox{\hskip1pt$\sim$}}
    \raise1pt\hbox{$>$}}}

\def\dslash{\not{\hbox{\kern-2pt $\partial$}}}
\def\Dslash{\not{\hbox{\kern-4pt $D$}}}
\def\Oslash{\not{\hbox{\kern-4pt $O$}}}
\def\Qslash{\not{\hbox{\kern-4pt $Q$}}}
\def\pslash{\not{\hbox{\kern-2.3pt $p$}}}
\def\kslash{\not{\hbox{\kern-2.3pt $k$}}}
\def\qslash{\not{\hbox{\kern-2.3pt $q$}}}

 \newtoks\slashfraction
 \slashfraction={.13}
 \def\slash#1{\setbox0\hbox{$ #1 $}
 \setbox0\hbox to \the\slashfraction\wd0{\hss \box0}/\box0 }

\def\ee{\end{equation}}
\def\be{\begin{equation}}

\begin{document}
\title{An improved limit on the neutrino mass with CMB and redshift-dependent
halo bias-mass relations from SDSS, DEEP2, and Lyman-Break Galaxies}

\author{Francesco De Bernardis$^{1,2}$,  Paolo Serra$^2$, Asantha Cooray$^2$, Alessandro Melchiorri$^1$}

\affiliation{$^1$Physics Department and sezione INFN, University
of Rome ``La Sapienza'', Ple Aldo Moro 2, 00185 Rome, Italy\\
$^2$Center for Cosmology, Dept. of Physics \& Astronomy, University of
California Irvine, Irvine, CA 92697}

\date{\today}%


\begin{abstract}

We use measurements of luminosity-dependent galaxy bias at several different
redshifts, SDSS at $z=0.05$, DEEP2 at $z=1$ and LBGs at $z=3.8$,
combined with WMAP five-year cosmic microwave background anisotropy data and SDSS Red Luminous
Galaxy survey three-dimensional clustering power spectrum to put constraints on
cosmological parameters. Fitting this combined dataset, we
show that the luminosity-dependent bias data that probe the relation between halo bias and halo mass and its redshift evolution
 are very sensitive to sum of the neutrino masses: in particular we obtain the upper limit of $\sum m_{\nu}<0.28$eV at
the $95\%$ confidence level for a $\Lambda CDM + m_{\nu}$ model, with a $\sigma_8$ equal to $\sigma_8=0.759\pm0.025$ (1$\sigma$).
When we allow the dark energy equation of state parameter $w$ to vary we find $w=-1.30\pm0.19$ for a
general $wCDM+m_{\nu}$ model with the 95\% confidence level upper limit on the neutrino masses at
$\sum m_{\nu}<0.59$eV. The constraint on the dark energy equation of state further improves to
$w=-1.125\pm0.092$ when using also ACBAR and supernovae Union data, in addition to above,
with a prior on the Hubble constant from the Hubble Space Telescope.
\end{abstract}
 \maketitle

\section{Introduction}

Galaxy clustering at large physical or angular scales corresponding
to the linear regime is a well-known probe of the cosmological
parameters. The combination of cosmic microwave background (CMB)
anisotropy data and the galaxy clustering spectrum, when combined
with additional probes such as distance measurements with Type Ia
supernovae or baryon acoustic oscillations, are known to break
various degeneracies between cosmological parameters that exist when
using either CMB data or galaxy clustering data alone
\cite{Tegmark:1997rp,Eisenstein:1998hr}.

While cosmological
parameters are generally derived from the shape of the galaxy power
spectrum by marginalizing over the overall uncertainty associated
with galaxy bias that relates clustering of dark matter to galaxies
at the linear scales, the galaxy bias, $b_g$, itself contains
certain cosmological information that is generally ignored. The
relation between galaxy bias and cosmology is evident in the context
of the halo model for galaxy clustering, which provides a simple way
to relate the galaxy distribution to that of the dark matter halo
distribution. The cosmological information is present in the
relation between dark matter halo distribution and the linear
density field and is captured by the dark matter halo bias as a
function of the halo mass $b_h(M)$ \cite{Mo}.  The dark matter halo bias
contains shape information of the power spectrum through rms
fluctuations of mass $\sigma(M)$.

In the halo model, either expressed in terms of a halo occupation
distribution (HOD; \cite{Abazajian:2004tn,CooShe,Scoccimarro:2000gm,Jing:1997nb,Cooray:2002uq,Zheng:2007zg,Zheng:2004id,Kravtsov:2003sg,Seljak:2000gq})
or a conditional luminosity function (CLF; \cite{Yang:2002ww,Yang:2004qi,Cooray:2005yt}),
one can relate the clustering bias of galaxies measured as a
function of the galaxy property, such as the luminosity $b_g(L)$, to
the  bias of dark matter haloes as a function of the halo mass. To derive the
relation between $b_g(L)$ and $b_h(M)$, it is necessary to have a proper
understanding of the relation between an observable quantity such as the galaxy
luminosity $L$ and a more fundamental quantity, the halo mass $M$ \cite{Cooray:2005yt,Cooray:2005mm,Cooray:2005qu,Yang:2003vq}.
This relation can be achieved through modeling of certain galaxy observables, such as the luminosity
function, non-linear or 1-halo part of the galaxy clustering
spectrum, and relations related to galaxy-mass observables from
galaxy-galaxy weak lensing measurements \cite{Cooray:2005ks,Vale:2004yt,More:2008yy}.
In the context of cosmological measurements, the $b_g(L)$ relation as measured by the
SDSS survey at low redshifts has been used to constrain cosmological
parameters, including the neutrino mass \cite{Seljak:2004sj,Seljak:2004xh}. When
combined with WMAP 1-year data, SDSS power spectrum shape, and the
SDSS $b_g(L)$ relation results in a 95\% confidence limit on the sum
of the neutrino masses of $\sim0.54$ eV.

Beyond SDSS, several galaxy surveys that target galaxy populations
at higher redshifts either through spectroscopic measurements or
through Lyman drop-out techniques have provided measurements related
to the galaxy luminosity functions and galaxy clustering power
spectra or correlation functions, as a function of the galaxy
luminosity. Among these surveys are the DEEP2 \cite{Coil:2003cf} at
$z \sim 1$ and Lyman-break galaxy surveys at $z \sim 3$ to 4 \cite{Kashikawa:2005xy,Ouchi:2003xw}. These
clustering and luminosity function measurements can be interpreted
in terms of a common CLF model \cite{Yang:2002ww,Cooray:2005yt,Cooray:2005mm} with which one can derive the
appropriate relation to connect galaxy luminosity $L$ to halo mass $M$ at a given redshift $z$ \cite{Cooray:2005ks,Vale:2004yt,More:2008yy}
terms of a conditional probability distribution function $P(M|L,z)$
\cite{Cooray:2005mm,Cooray:2005qu,Cooray:2006aq}.

Given that clustering measurements at large linear scales lead to estimates of $b_g(L)$ at
a redshift different from SDSS and we also have a mechanism to
connect $b_g(L)$ to $b_h(M)$ through a statistical description, this
raises the possibility of further constraining the cosmological
parameters than using SDSS galaxy power spectrum and SDSS galaxy
bias-luminosity relation alone. In addition to the shape information
captured by $b_g(L)$ at each redshift, the overall evolution of
$b_g(L)$ as a function of redshift is further sensitive to the
linear growth function of dark matter fluctuations. Since the growth
function depends strongly on properties of the dark energy, such as
the equation-of-state (EOS) relating the ratio of dark energy
pressure to density, the combination of $b_g(L)$ measurements at
several redshifts raises the possibility of constraining the EOS, in
addition to cosmological parameters that probe the shape of the dark
matter power spectrum.

The paper is organized as follows. In \S~2 we provide a brief
summary on how cosmological information can be extracted from
$b_g(L,z)$ measurements by making use of the relations between
galaxy luminosity and halo mass, captured  by the probability distribution of a galaxy with a luminosity $L$ to appear in a halo mass of mass $M$ at a redshift $z$, $P(L|M,z)$,
from CLF modeling described in Ref.~\cite{Cooray:2005mm,Cooray:2006aq}.
In \S~3, we describe the analysis of all data. In
addition to $b_g(L)$ measurements at three different redshifts, we
also make use of WMAP 5-year data\cite{Komatsu:2008hk} (by updating WMAP first-year data used
in the analysis of \cite{Seljak:2004sj}) and shape of the SDSS Luminous Red
Galaxy (LRG) power spectrum \cite{Tegmark:2006az} at low redshifts. \S~4 presents
our results and we concluded with a summary of important constraints
on cosmological parameters in \S~5.

\section{Modelling the Bias}

The clustering of bound viralized objects is biased with respect to that of
underlying dark matter distribution  and with the decreasing number density of the objects,
the bias factor is known to increase \cite{Mo}.
Thus, bright galaxies that are in rare massive halos are expected to be more biased
that less luminous and abundant galaxies. Here, we model the relation between galaxy bias
and halo bias following an approach similar to that of Ref.~\cite{Seljak:2004sj},
 but using the improved halo bias relation from Ref.~\cite{Sheth:1999su}
 corresponding to the ellipsoidal collapse model instead of the fitting function for bias \cite{Seljak:2004ni}.
We also generalize this relation to higher redshifts (see, Appendix A of Ref.~\cite{Tinker:2004gf}):
\begin{equation}\label{sht}
\begin{split}
b_h(\nu(z))=1+\frac{1}{\sqrt{a}\delta_c}[\sqrt{a}(a\nu^2)+\\\sqrt{a}b(a\nu^2)^{1-c}-\frac{(a\nu^2)^c}{(a\nu^2)^c+b(1-c)(1-c/2)}] \, ,
\end{split}
\end{equation}
in this expression $\delta_c=1.686$ is the threshold overdensity
required for collapse of an over-density region and
$\nu(z)=\delta_c/\sigma(M,z)$. The parameters $a$,$b$ e $c$ are
constants and we use the values suggested in Ref.~\cite{Sheth:1999su} with
$a=0.707$, $b=0.5$, and $c=0.6$. The quantity $\sigma(M,z)$ is the rms
mass fluctuation in spheres with radius $r=(3M/4\pi\bar{\rho})^{1/3}$,
where $M$ is the halo mass and $\rho$ the mean matter density at
redshift $z$. $\sigma(M,z)$ can be calculated through the relation:
\begin{equation}\label{sig}
    \sigma^2(M,z)=\frac{1}{2\pi^2}\int P(k,z)W^2(k)k^2dk
\end{equation}
where $W(k)$ is the Fourier transform of ttop-hat window function.
To compute equation~(\ref{sig}) we use the linear matter power spectrum
$P(k,z)$ generated by CAMB at redshift z for a given set of cosmological parameters.
The dependence of bias on cosmological parameters is contained in the
quantity $\nu(z)=\delta_c/\sigma(M,z)$ through
information from the linear matter power spectrum from $\sigma(M,z)$.

As shown in several works involving measurements with data
galaxy bias, as measured from the galaxy power spectrum at large physical scales corresponding to linear
regime of clustering depends on luminosity, with brighter
galaxies more strongly clustered than fainter ones \cite{Norberg:2001wh,Tegmark:2003uf}.
In the halo model, galaxies are expected to populate dark matter halos and these halos are already biased with respect to the density field
$b_h(M,z)$, where $M$ is the mass of a halo at a redshift $z$. This is the quantity that is directly linked to cosmology, while bias measurements
directly from data are a function of luminosity. We can relate the two through the
probability distribution $P(M;L,z)$ \cite{Guzik:2002zp,Yang:2003vq,Cooray:2005mm}
that a galaxy of luminosity  $L$ resides in a halo of mass $M$.

If we know the $P(M;L,z)$ then the bias
at a fixed luminosity is given by:
\begin{equation}\label{biaslum}
    b(L,z)=\int P(M;L,z)b_h(M,z)dM
\end{equation}
\begin{figure}[h]
 \includegraphics[width=188pt]{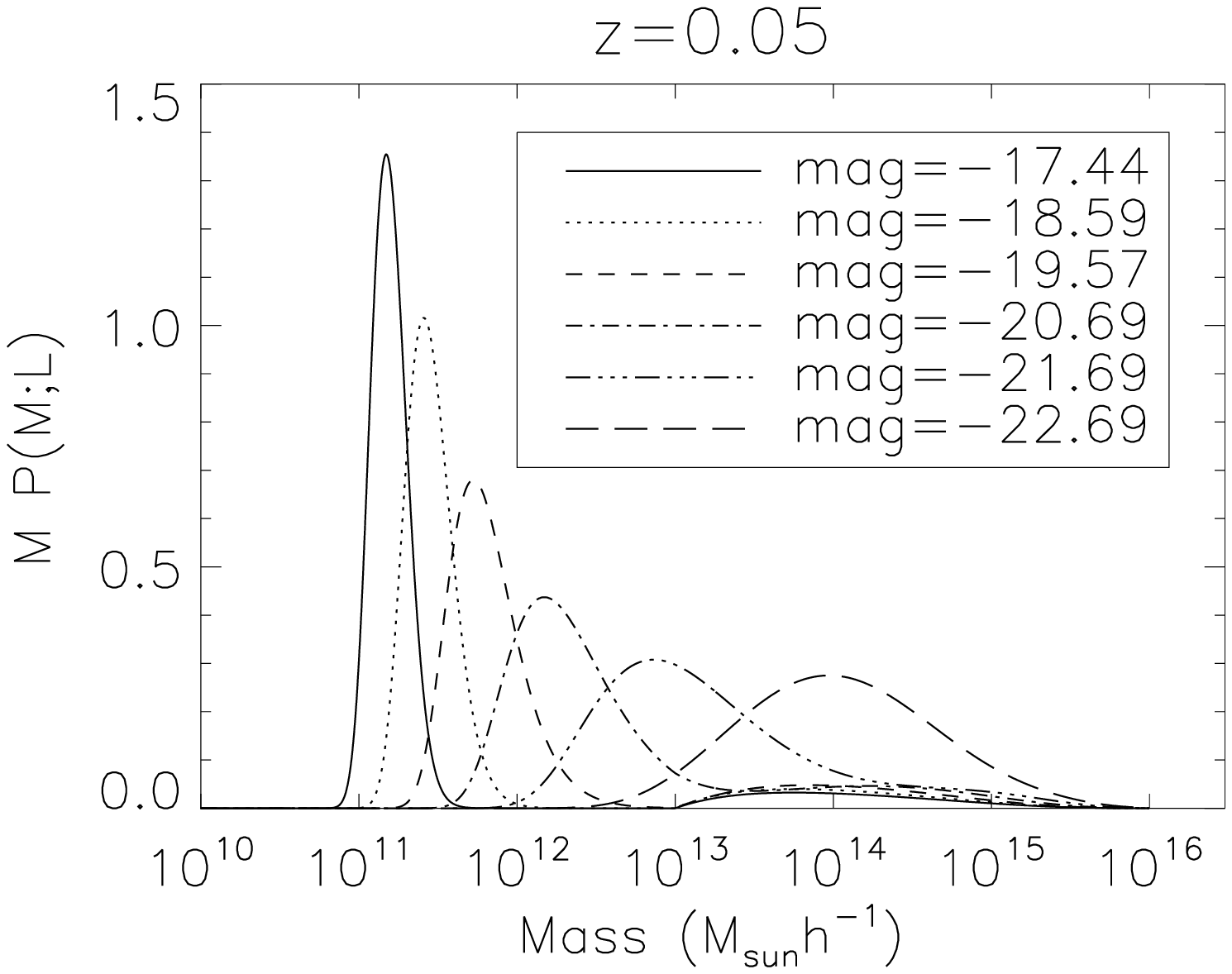}\\
 \includegraphics[width=188pt]{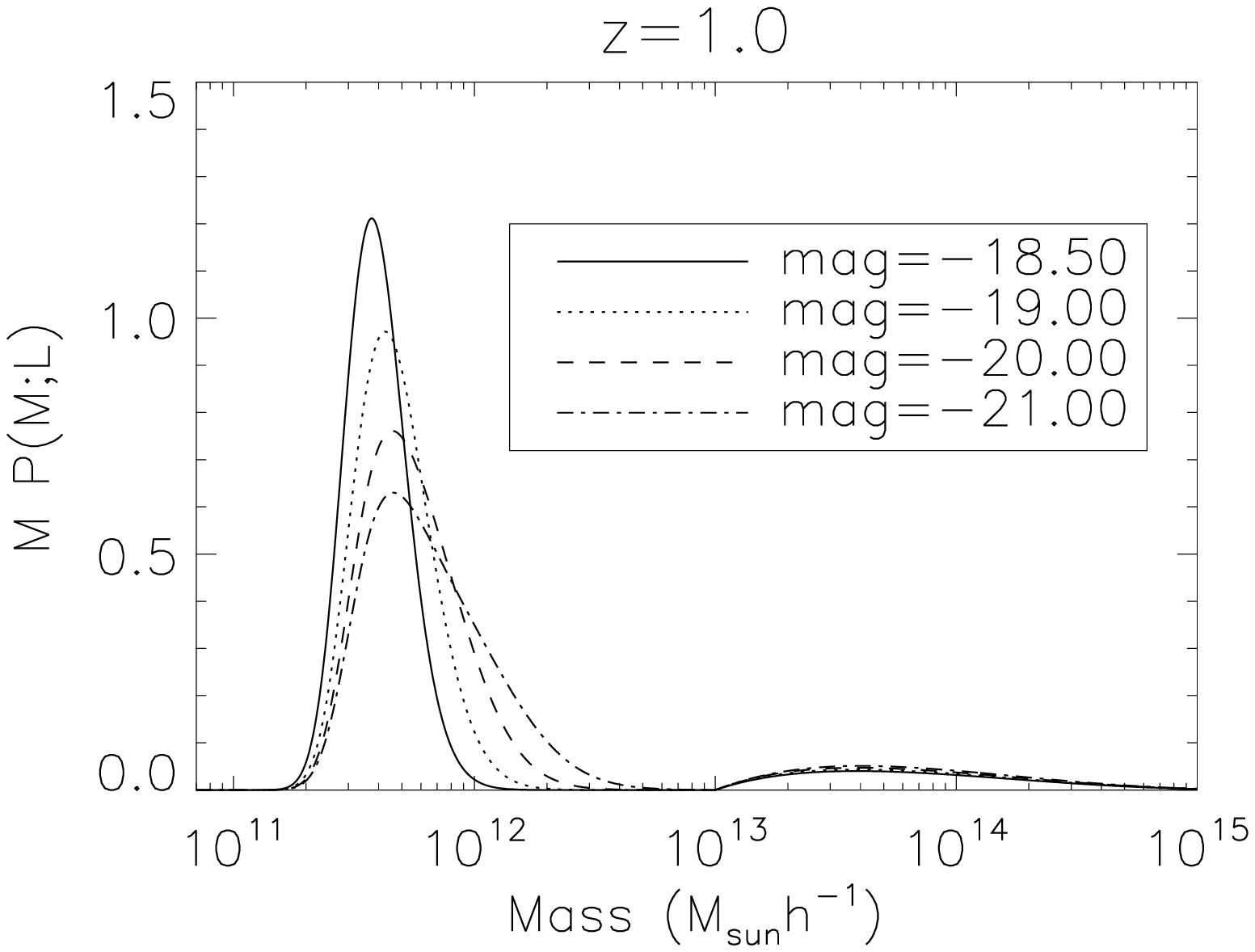}\\
 \includegraphics[width=188pt]{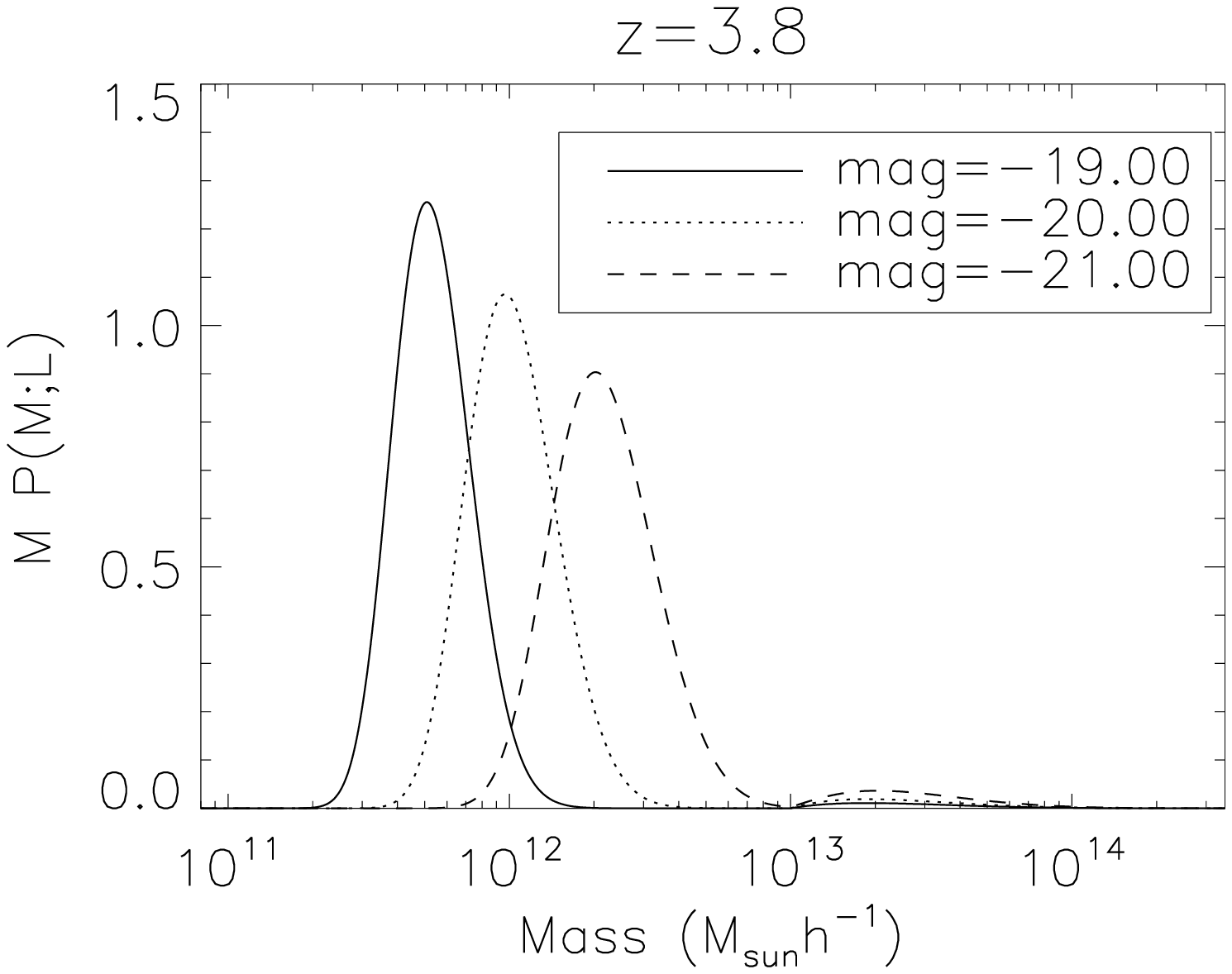}\\
\caption{The conditional probability distribution $P(M;L,z)$ relating the galaxy luminosity $L$ and halo mass $M$, at different redshifts, as calculated in
Refs.~\cite{Cooray:2005mm,Cooray:2006aq} for
SDSS at $z \sim 0.05$, DEEP2 at $z \sim 1$, and LBGs at $z \sim 4$. The probabilities to find a galaxy at a given luminosity in a halo of mass $M$ at redshift $z$
is plotted  as a function of the halo mass for luminosity values for which we have galaxy bias data.
In each of the distributions, the peak at low halo masses is related to galaxies of the given luminosity that appear as central galaxies, while the tail
extending to higher masses is for galaxies that appear as satellites in more massive halos. The width of the central peak is related to the scatter in the
relation between  luminosity of central galaxies and halo mass and cannot simply be described by a delta function relating a one-to-one correspondence between mass and luminosity
\cite{Cooray:2005ks,Vale:2004yt,More:2008yy}.}
\label{pd}
\end{figure}

For SDSS galaxies, the conditional probability $P(M;L,z)$ at low redshifts was derived based on a combination
if SDSS galaxy luminosity function \cite{Blanton:2004zy} and luminosity-dependent galaxy correlation functions \cite{Zehavi:2004ii} that probe the
non-linear, 1-halo term of the halo model \cite{CooShe}. The luminosity function is a strong probe of the $L_c(M)$ relation relating the luminosity of central galaxies to their host dark matter halo mass,
as well as an average scatter in that relation \cite{Cooray:2005yt}, while the non-linear (1-halo) part of the galaxy clustering, either the correlation function or the power spectrum,
establishes information related to the CLF of satellite galaxies.  The large, linear scale clustering provides necessary information related to $b_g(L)$.
The degeneracies in the model parameters related to the CLF parameterization is broken with additional data such as the of galaxy-mass correlation function from
SDSS galaxy-galaxy lensing measurements, similar to the analysis in Ref.~\cite{Seljak:2004sj}, and we make use of results from
publicly available SDSS galaxy-mass correlation functions \cite{McKay:2001gt,Mandelbaum:2004mq} in Ref.~\cite{Cooray:2005mm}.
In the case of DEEP2  and higher redshift LBG data, we again use galaxy clustering (DEEP2: Ref.~\cite{Coil:2003cf}; Subaru LBG: Ref.~\cite{Ouchi:2003xw} and luminosity functions measurements (DEEP2: Ref.~\cite{Willmer:2005fk}; LBG: Ref.~\cite{Kashikawa:2005xy}) from the literature \cite{Cooray:2006aq}.

These conditional probability distributions are plotted in fig.\ref{pd} as a function of redshift and luminosity.
The distribution functions account for both central galaxies and the satellite galaxies, following the conditional luminosity function
approach of Refs.~\cite{Cooray:2005mm,Cooray:2006aq}.
Due to scatter in the relation between luminosity of central galaxies and the halo mass, the distributions have a scatter even for the
central galaxy peak at the low-mass end. The previous analysis in Ref.~\cite{Seljak:2004sj} ignored this scatter and described the relation between
central galaxy luminosity and halo mass with a delta function and assumed simple model description with one free parameter to describe the same relation
for satellites. The distributions shown in fig.~\ref{pd} have additional uncertainties due to limitations in constructing CLFs and when
fitting to data, we allow for this uncertainty in two ways: to account for an overall systematic error, we marginalize over a nuisance parameter $b_*$
that scales the bias values by an overall  factor and we include an additional error in bias measurements.

\begin{figure}[h]
  \includegraphics[width=260pt]{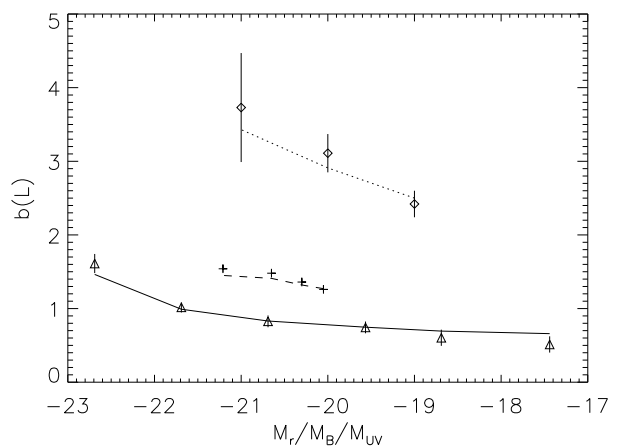}
  \caption{The galaxy bias-luminosity data set used in our analysis for the three average redshifts of SDSS ($z \sim 0.05$), DEEP2 ($z\sim 1$), and LBG ($z\sim 3.8$)
in comparison with the bias prediction calculated for the best fit $\Lambda$CDM
  model. The x-axis magnitude values plotted are $M_r$ for SDSS, $M_B$ for DEEP2, and $M_{\rm UV}$
for LBGs at $z\sim 3.8$.}
\label{data}
\end{figure}

\begin{figure*}[t]
  \begin{center}
    \begin{tabular}{ccc}
      \resizebox{50mm}{!}{\includegraphics{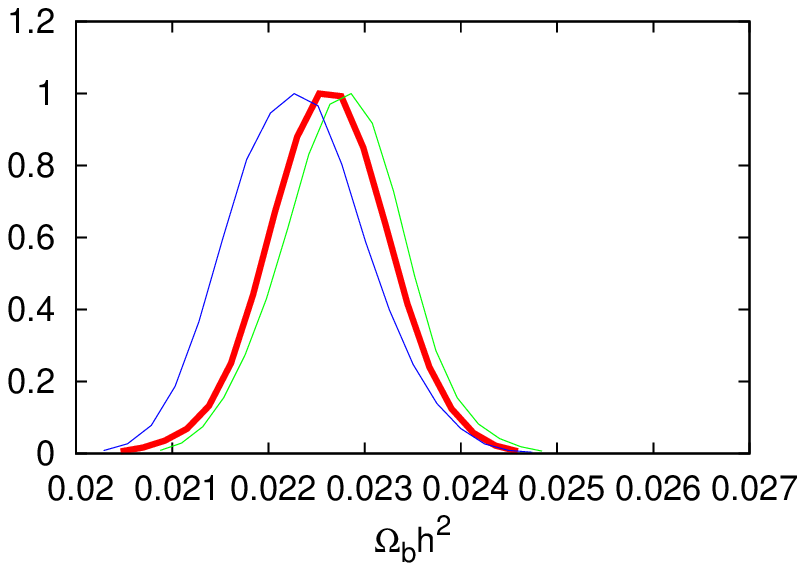}} &
      \resizebox{50mm}{!}{\includegraphics{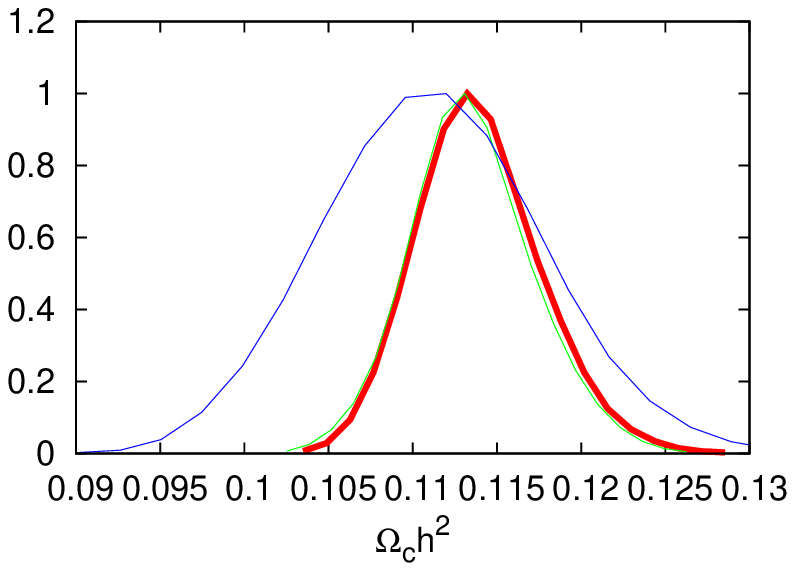}} &
      \resizebox{50mm}{!}{\includegraphics{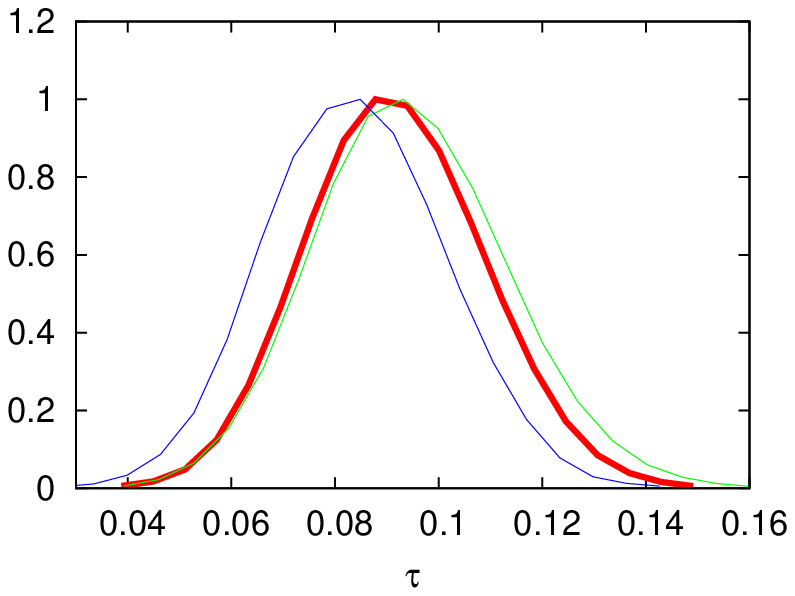}}  \\
      \resizebox{50mm}{!}{\includegraphics{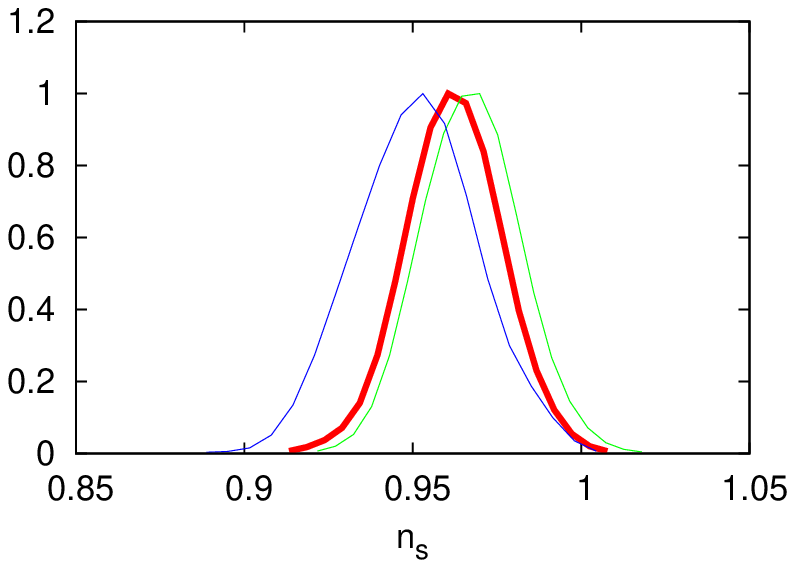}} &
      \resizebox{50mm}{!}{\includegraphics{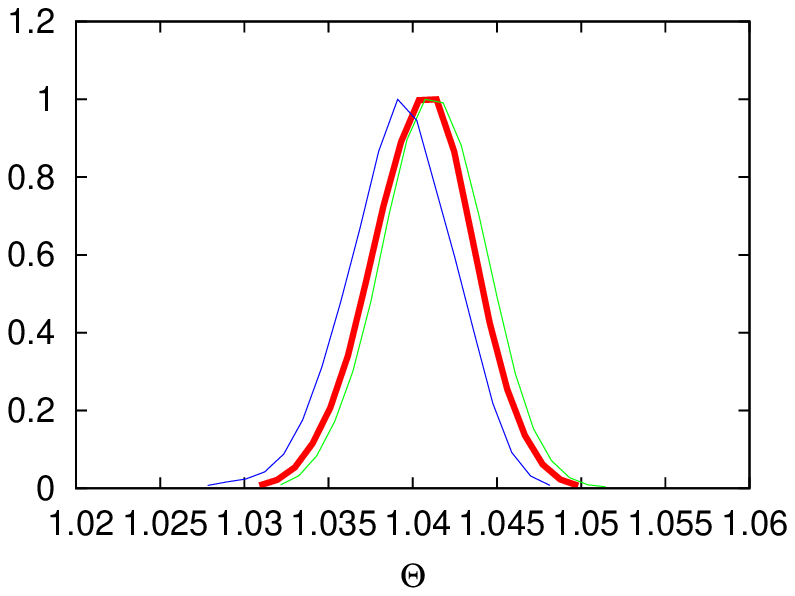}} &
      \resizebox{50mm}{!}{\includegraphics{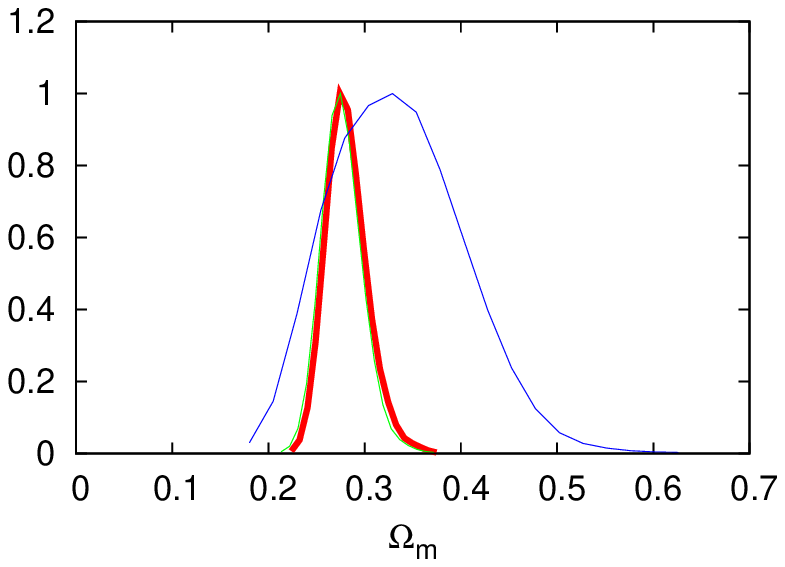}} \\
      \resizebox{50mm}{!}{\includegraphics{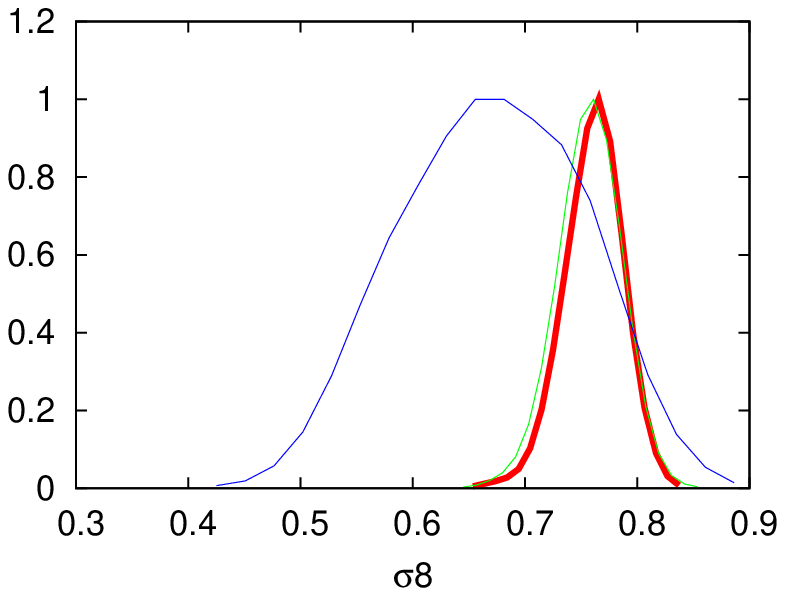}} &
      \resizebox{50mm}{!}{\includegraphics{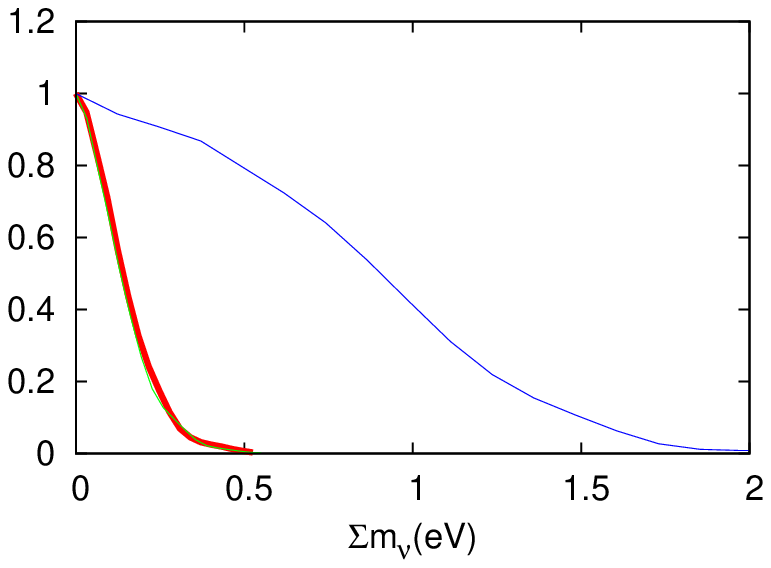}} &
      \resizebox{50mm}{!}{\includegraphics{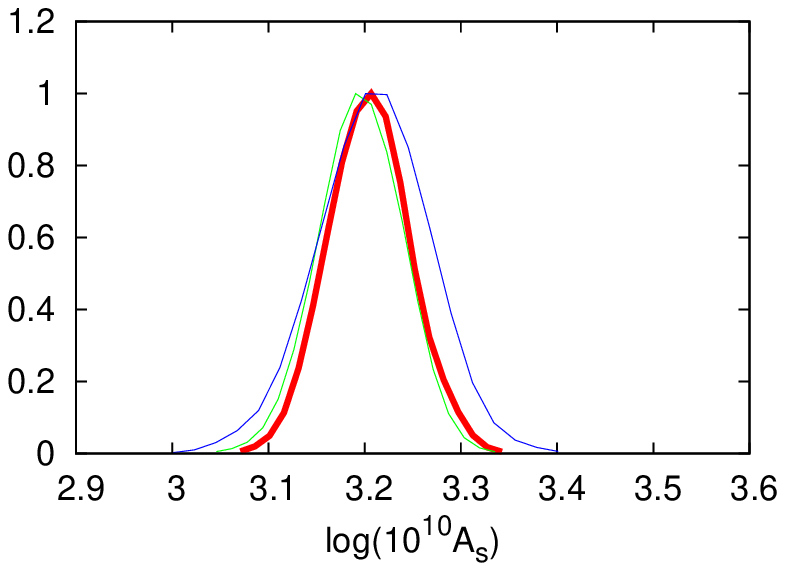}} \\
    \end{tabular}
    \caption{constraints on the parameters of the $\Lambda$CDM+$m_{\nu}$ model from WMAP alone (blue), WMAP+LRG+bias data set at z=0.05
(green) and WMAP+LRG+all bias data sets (red).}
    \label{test4}
  \end{center}
\end{figure*}

\section{Analysis}

To fit bias measurements together with CMB and SDSS we performed a
Monte Carlo Markov Chain analysis using a modified version of the
publicly available code cosmoMC \cite{Lewis:2002ah}, with a
convergence diagnostic based on the Gelman and Rubin statistic
\cite{GR} (also known as ``R-1'' statistic, where $R$ is defined as the
parameter R is defined as the ratio between the
variance of chain means and the mean of variances).
Our cosmoMC runs consist of 2-3 chains  typically
with 15000-20000 points and we have for our
chains $R-1<0.01$, ignoring first $50\%$ of the chains.

For each cosmological model we repeated the procedure
described in the previous Section to calculate the theoretical $b_g(L)$ relation and
implemented, at each redshift, a relation similar to that used in
Ref.~\cite{Seljak:2004sj} to compare with data:
\begin{equation}\label{chiq}
    \chi^2=\sum_i\frac{(b_{th,i}-b_*(b_{data}/b_*)_i)^2}{b^2_*\sigma^2_{b/b_{*,i}}+\sigma^2_{sys}}
\end{equation}
where $b_{th,i}$ is the predicted bias at given luminosity for a given cosmological model,
$b_{data}/b_*$ is the observed bias at the same luminosity with
error $\sigma_{b/b_{*,i}}$, and $\sigma_{sys}=0.03$ is a systematic
uncertainty in the modeling of bias \cite{Seljak:2004ni}. The sum
is over the number of bias data points at each redshift. $b_*(z)$ is the
bias parameter, as a function of redshift, that accounts for an overall uncertainty in the
bias measurements or modeling of bias based on CLFs. We treat it as a free
parameter and marginalize over it when quoting cosmological parameter errors.  This parameter shifts the model (or data)
by a constant factor while keeping the shape the same. Thus, cosmology is measured through the shape of the $b_g(L)$ relation
and not from its exact amplitude.

We included equation~(\ref{chiq}) into the likelihood for the
five-year WMAP data \cite{Komatsu:2008hk} and SDSS LRG power spectrum \cite{Tegmark:2006az}.
We sample first the following simple seven-parameters cosmological model assuming
flat priors on parameters and treating the dark energy component as
a cosmological constant: the physical baryon and cold dark matter
densities, $\Omega_bh^2$ and $\Omega_ch^2$, the ratio of sound
horizon to the angular diameter distance at decoupling, $\theta_s$,
the overall normalization of the spectrum at $k=0.002$ h Mpc$^{-1}$,
$A_S$, the amplitude of SZ spectrum, $A_{SZ}$, the optical depth to
reionization $\tau$, and the scalar spectral index $n_s$.
In our analysis we always assume spatial flatness ($\Omega_k=0$). The bias
parameter depends on redshift, so when using all redshift data set
we introduce three free bias parameters for $b_*$ in our analysis in addition
to the cosmological parameters listed above, for a total of ten free parameters.

We also explored a larger set of parameters, introducing the sum
of neutrino masses $\sum m_{\nu}$ and the dark energy equation of
state $w$. When including both neutrino masses and $w$ we performed
an analysis combining bias data with  WMAP and LRG only and one also
using Arcminute Cosmology Bolometer Array Receiver (ACBAR) data
\cite{Reichardt:2008ay} and luminosity distance SN-Ia data (SNe)
\cite{Astier:2005qq,Riess:2004nr} assuming the prior from
Hubble Space Telescope (HST) on the value of Hubble constant
$h=0.72\pm0.07$ \cite{Freedman:2000cf}.

We use galaxy bias data at three different redshifts: from Sloan
Digital Sky Survey \cite{Tegmark:2003uf}, six points between
redshift $0.05$ and $0.1$, from DEEP2 redshift survey
\cite{Coil:2003cf} four points at $z\sim0.8$ to $z \sim 1.1$, and
from clustering of Lyman Break Galaxies in the Subaru Deep Field
\cite{Ouchi:2003xw,Cooray:2006aq} with three points at $z=3.8$.
These data points are shown in fig.~\ref{data} with the best fit
model.

\begin{figure}
  \begin{center}
  \includegraphics[width=250pt]{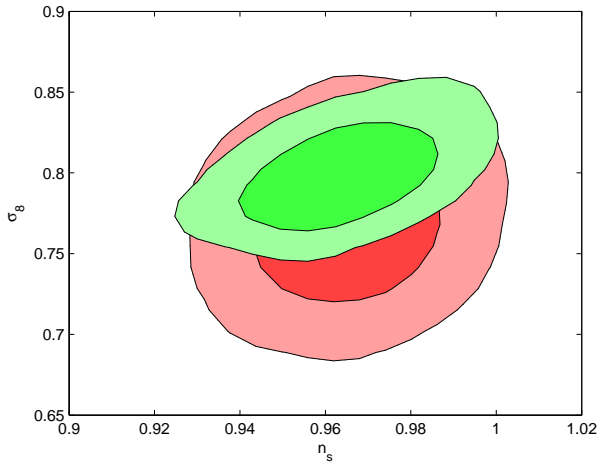}\\
  \caption{joint two-dimensional posterior probability contour plot in the $\sigma_8$-$n_s$ plane showing $68\%$ and
  $95\%$ contours from WMAP alone (red) and WMAP+LRG+bias data at all redshifts (green). }\label{s8ns}
  \end{center}

\end{figure}

\begin{figure}
 \begin{center}
  \includegraphics[width=250pt]{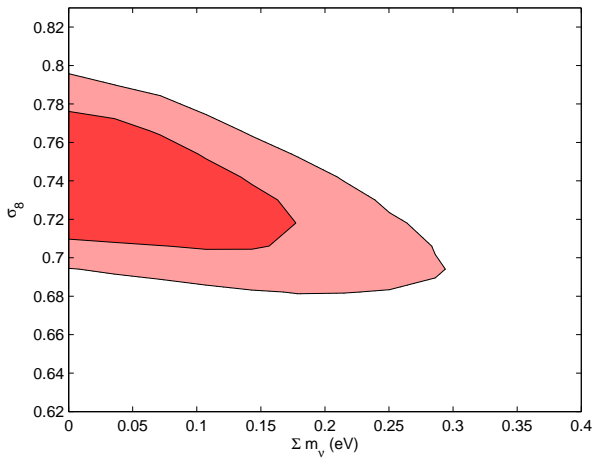}\\
  \caption{joint two-dimensional posterior probability contour plot in the $\sigma_8$-$\sum m_{\nu}$ plane showing $68\%$ and
  $95\%$ contours from WMAP+LRG+bias data at all redshifts.}\label{s8mnu}
  \end{center}
\end{figure}

\begin{figure}
  \begin{center}
  \includegraphics[width=250pt]{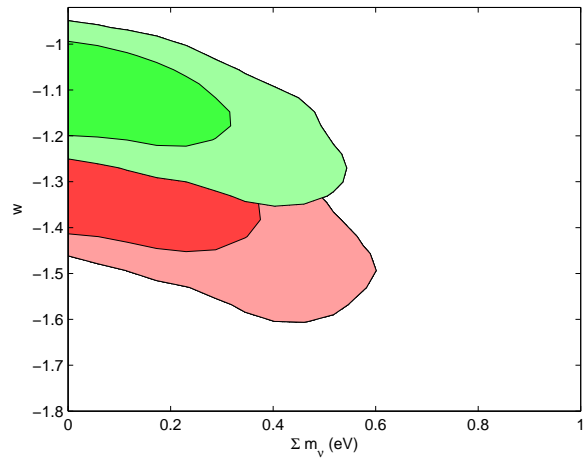}\\
  \caption{joint two-dimensional posterior probability contour plot in the $\sum m_{\nu}$-$w$ plane showing $68\%$ and
  $95\%$ contours from WMAP+LRG and bias data at all redshifts (red) and
WMAP+ACBAR+SNe+LRG+HST and bias data at all redshifts (green). }\label{wmnu}
  \end{center}
\end{figure}

\begin{table*}
\begin{center}
\begin{tabular}{lrrr|rrr}
\hline
&&$\Lambda$CDM+$w$&&&$\Lambda$CDM+$w+m_{\nu}$&\\
&WMAP5 &&+SDSS+all z's&WMAP5&&+SDSS+all z's \\
&        &     \\
\hline \hline
 $ \Omega_b h^2$ & $0.02273\pm0.00065$   & & $0.02248\pm0.00060$    &    $0.02222\pm0.00063$  &&    $0.02234\pm0.00060$    \\
 $ \Omega_c h^2$ & $0.1102\pm0.0065$      && $0.1160\pm0.0041$      &    $0.1119\pm0.0064$    &&    $0.1223\pm0.0065$    \\
 $ \tau$  & $0.086\pm0.017$               & & $0.087\pm0.017$       &    $0.083\pm0.016$      &&    $0.083\pm0.016$    \\
 $ n_s$ & $0.963\pm0.016$                 && $0.965\pm0.014$        &    $0.948\pm0.017$      &&    $0.954\pm0.014$    \\
 $w$    & $-1.06\pm0.41$                  &&  $-1.12\pm0.10$        &    $-1.23\pm0.55$       &&    $-1.30\pm0.19$    \\
 $ln(10^{10} A_s)$     &   $3.18\pm0.05$        && $3.22\pm0.04$   &             $3.22\pm0.06$              && $3.24\pm0.05$\\
 $ \Omega_m$ &    $0.27\pm0.10$           && $0.259\pm0.020$        &    $0.31\pm0.13$        &&    $0.267\pm0.027$    \\
 $\sigma_8$ &   $0.81\pm0.14$             && $0.802\pm0.037$        &    $0.71\pm0.14$        &&    $0.775\pm0.045$    \\
 $\sum m_{\nu}$    &  $-$                 && $-$                    &    $<1.5$eV  (95\%CL)   &&    $<0.59$eV (95\%CL)\\
 $ b^*_1$& $-$                            && $1.01\pm0.03$          &    $-$                  &&    $1.03\pm0.04$    \\
 $ b^*_2$ & $-$                           & & $1.18\pm0.06$         &    $-$                  &&    $1.21\pm0.07$   \\
 $ b^*_3$ & $-$                           & & $3.44\pm0.35$         &    $-$                  &&    $3.44\pm0.35$    \\
 \hline
\hline
\end{tabular}
\caption{Mean values and $1\sigma$ constraints on cosmological
parameters from WMAP+SDSS+bias data at all redshifts in comparison
with constraints from WMAP alone, for models with dark energy
equation of state allowed to vary.}\label{table:2}\vspace{1cm}
\end{center}
\end{table*}

\section{Results}

The constraints on cosmological parameters are shown in the
Tables~\ref{table:2}, \ref{table:1} and \ref{table:3} with a
comparison to constraints from WMAP five-year data
\cite{Komatsu:2008hk} alone both for a simple $\Lambda$CDM model
(Table~\ref{table:1}) and for a model with a non-zero mass for
neutrinos ($\Lambda$CDM+$m_\nu$) (Table~\ref{table:2}), and with a
dark energy equation of state different from the cosmological
constant value of $-1$ in addition to neutrino mass
($\Lambda$CDM+$m_\nu$+$w$) (Table~\ref{table:3}). In addition to
WMAP data, we also consider the combination of WMAP data and SDSS
LRG power spectrum shape with $b_g(L)$ relation from SDSS, and
finally the same data complemented with high-redshift $b_g(L)$
relations from DEEP2 and Subaru LBGs.

As we have discussed in the introduction galaxy bias depends on rms fluctuation $\sigma(M)$ in spheres that contains a
mass $M$. Galaxy bias measurements, as a function of luminosity, are therefore able to constrain all
cosmological parameters that affect this quantity, mainly  the
amplitude of matter fluctuations $\sigma_8$, power spectrum spectral index or tilt $n_s$, and neutrino mass, that
affect the growth of density perturbations.  As presented in Ref.~\cite{Seljak:2004sj}, bias data at low redshifts from SDSS are
already strongly sensitive to neutrino masses:  with WMAP first-year data combined with SDSS galaxy power spectrum shape and
SDSS $b_g(L)$ data lead to $\sum m_{\nu} < 0.54$ eV at the 95\% confidence level (See Table~\ref{table:res} for recent results on neutrino masses).

With WMAP 5-year data and SDSS LRG power spectrum complemented by galaxy bias data at $z\sim 0.05$, 1, and 3.8, we are able to
improve constraints on the sum of neutrino mass by a factor $\sim2$ with
respect to the result of Ref.~\cite{Seljak:2004sj} obtaining $\sum m_{\nu}<0.28$ eV at the 95\% confidence level.
We get a similar result if we only keep to $z \sim 0.05$ SDSS $b_g(L)$ data and the LRG power spectrum shape with WMAP 5-year data,
since by adding additional bias data at higher redshifts we are also introducing to the analysis two more unknown parameters, i.e. the nuisance
bias normalization parameters for redshifts $z=1$ and $z=3.8$, which are marginalized over when quoting parameter errors.
The relative increase of a factor of $\sim 2$ in the neutrino mass limit
compared to Ref.~\cite{Seljak:2004sj} is part due to the improvement in both the CMB (WMAP one-year to WMAP five-year)
and galaxy power spectrum shape data (SDSS DR2 power spectrum with $\sim 200,000$ galaxies to SDSS DR4 LRG power spectrum with $\sim 400,000$ galaxies)
ad part due to the improvement in the CLF modeling of the $P(M|L,z)$ relation for SDSS galaxies.
While the combination of all bias data at the three redshifts does not improve the limit on the sum of neutrino masses
compared to the case with bias measurements from SDSS only, we do find small improvements in the uncertainties of the other parameters,
as shown in fig.~\ref{test4} for the case of the $\Lambda$CDM model with a non-zero mass for neutrinos.
We plot also probability contours in $\sigma_8$-$n_s$ plane and in $\sum m_{\nu}$-$\sigma_8$
plane in figs.~\ref{s8ns} and \ref{s8mnu}.

As the growth of structure depends also on the dark energy density and equation of state we explored a more
general parameter space, relaxing the assumption of a cosmological
constant for dark energy and constraining the equation of state of dark energy $w$
both in the case of neutrino mass fixed to zero and allowed to vary (Table~\ref{table:2}). We find that $w=-1.06 \pm 0.41$ with WMAP 5-year data alone
and $-1.12 \pm 0.10$ with WMAP 5-year+SDSS LRG power spectrum shape and all $b_g(L)$ data. For comparison, the WMAP 5-year data
combined with Baryon Acoustic Oscillation (BAO) data \cite{Percival:2007yw} gives $w=-1.15 \pm 0.21$.

Our constraints on $\sigma_8$ are $\sigma_8=0.759\pm0.025$ in the
case $\Lambda CDM+m_{\nu}$, while WMAP combined BAOs \cite{Percival:2007yw} and
SNe data gives $\sigma_8=0.732\pm0.062$. The analysis of bias
data combined with CMB and SDSS showed in \cite{Seljak:2004sj} gives
$0.854\pm0.062$.

For the most general parameter space explored in our analysis, with
both sum of the neutrino masses and dark energy equation of state
allowed to vary, we improve constraints on equation of state with
respect to WMAP alone ($w=-1.23\pm0.54$), obtaining
$w=-1.30\pm0.19$. When $w$ is allowed to vary, constraints on
neutrino masses are weakened to $\sum m_{\nu}<0.59$ eV at the $95\%$
confidence level, but are still improved with respect to $\sum
m_{\nu}<0.66$eV at the same $95\%$ confidence level  with
WMAP+BAO+SNe, for the same model. We then performed the same
analysis combining WMAP, LRG and all bias data set with ACBAR and
SNe data, assuming the prior from Hubble Space Telescope (HST) on
the value of Hubble constant and we improve the last constraints on
neutrino masses and $w$, obtaining $\sum m_{\nu}<0.56$eV at the
$95\%$ confidence level and $w=-1.125\pm0.092$. These results are
shown in table \ref{table:3}. In fig \ref{wmnu} we plot the
probability contours for $w$ and $\sum m_{\nu}$  from WMAP combined
with LRG and bias data sets and the same combined also with ACBAR
and SNe, with the prior from HST. In the last case the inclusions of
SNe and ACBAR data and of the prior on $h$ slightly improves the
constraints on $\sum m_{\nu}$ but above all improves constraints on
$w$ (see also tables \ref{table:2}-\ref{table:3}).

\begin{table*}
\begin{center}
\begin{tabular}{lrrr|rrrrrrrrrrrr}
\hline
&&$\Lambda$CDM&&&$\Lambda$CDM+$m_{\nu}$&&\\
&WMAP5 &&          +SDSS+all z's& WMAP5 &&          +SDSS+all z's \\
&        &  &   \\
\hline \hline
 $ \Omega_b h^2$ & $0.02273\pm0.00062$  &  & $0.02266\pm0.00057$ &  $0.02226\pm0.00063$& & $0.02267\pm0.00058$   \\
 $ \Omega_c h^2$ & $0.1099\pm0.0062$     &  &    $0.1131\pm0.0034$& $0.1110\pm0.0062$  & & $0.1141\pm0.0038$   \\
 $ \tau$ & $0.087\pm0.017$ &             & $0.092\pm0.017$        & $0.084\pm0.016$  &   &  $0.094\pm0.016$     \\
 $ n_s$ & $0.963\pm0.014$            &   &            $0.964\pm0.013$ &  $0.950\pm0.017$ && $0.964\pm0.013$\\
 $ln(10^{10} A_s)$      &  $3.18\pm0.05$  &                & $3.20\pm0.04$ & $3.21\pm0.05$& & $3.21\pm0.04$\\
 $ \Omega_m$ &    $0.258\pm0.030$          & &          $0.273\pm0.017$     & $0.331\pm0.066$& & $0.282\pm0.023$ \\
 $\sigma_8$ &   $0.796\pm0.036$             & &        $0.807\pm0.021$ &  $0.675\pm0.084$  && $0.759\pm0.025$ \\
 $\Sigma m_{\nu}$  & $-$ && $-$ &       $<1.3$eV (95\%CL)   &               &     $<0.28$eV (95\%CL)\\
 $ b^*_1$ & $-$ &                        &  $1.01\pm0.03$       &  $-$  &   & $1.03\pm0.03$\\
$ b^*_2$ & $-$ &                         &         $1.21\pm0.06$  &  $-$ & &$1.24\pm0.06$\\
$ b^*_3$ & $-$ &                         &            $3.44\pm0.34$  & $-$& & $3.41\pm0.33$    \\
\hline \hline
\end{tabular}\vspace{1cm}
\begin{tabular}{lr|r|r}
\hline
 \end{tabular}
\caption{mean values and $1\sigma$ constraints on cosmological
parameters from WMAP+SDSS+bias data at all redshifts in comparison
with constraints from WMAP alone, for $\Lambda$ CDM and $\Lambda
$CDM+$m_{\nu}$ models.}\label{table:1}
\end{center}
\end{table*}
\begin{table*}
\begin{tabular}{lll}
  \hline \hline
  Kahniashvili et al. (2005) \cite{Kahniashvili:2005sg}  & Cluster Number Density & $\sum m_{\nu}<2.4$ eV\\
  Komatsu et al. (2008) \cite{Komatsu:2008hk} & WMAP5  &  $\sum m_{\nu}<1.3$ eV \\
  Tegmark et al. (2006)    \cite{Tegmark:2006az} & SDSS+WMAP3 &         $\sum m_{\nu}<0.9$ eV\\
  Komatsu et al. (2008) \cite{Komatsu:2008hk} & WMAP5+BAO+SNe & $\sum m_{\nu}<0.61$ eV  \\
  Kristiansen et al. (2007) \cite{Kristiansen:2007rx}  & WMAP3+SDSS+SNLS+BAO+CMF &$\sum m_{\nu}<0.56$ eV\\
  Seljak et al. (2005) \cite{Seljak:2004sj} & WMAP1+SDSS+SDSS $b_g(L)$ & $\sum m_{\nu} < 0.54$ eV\\
  Mac Tavish et al. (2005) \cite{MacTavish:2005yk}  & CMB+LSS &    $\sum m_{\nu}<0.48$ eV \\
  Seljak et al.     (2004) \cite{Seljak:2004xh}  & WMAP1+SDSS+Ly$\alpha$   & $\sum m_{\nu}<0.42$ eV \\
  Kristiansen et al. (2007) \cite{Kristiansen:2007rx}  & WMAP3+SDSS+SNLS+BAO+HST &$\sum m_{\nu}<0.40$ eV\\
  Fogli et al. (2008)   \cite{Fogli:2008ig}            & CMB+HST+SNe+BAO+Ly-$\alpha$& $\sum m_{\nu}<0.19$ eV\\
  Seljak et al. (2006)    \cite{Seljak:2006bg}        & CMB+SDSS+2dF+SNe+Ly-$\alpha$ &$\sum m_{\nu}<0.17$ eV\\
\hline
This paper                                            & WMAP5+SDSS LRG+ SDSS, DEEP2, LBG $b_g(L)$ & $\sum m_{\nu}<0.28$ eV\\
  \hline \hline
\end{tabular}\caption{Summary of the constraints at $95\%$ confidence level on the sum of neutrino masses from various data sets in the literature.
CMB means the collection of CMB data sets listed respectively in
\cite{Fogli:2008ig}, \cite{Seljak:2006bg} and
\cite{MacTavish:2005yk}. LSS means combination of SDSS and 2dF data
sets. SNLS is the Supernova Legacy Survey, CMF is the cluster mass
function, and Ly-$\alpha$ are the clustering measurements of the
Lyman-$\alpha$ forest. We refer the reader to individual references
given in the table for additional details related to datasets used
and how constraints were dervied.} \label{table:res}
\end{table*}

It is interesting to note that these limits on the sum of neutrino masses are
almost the strongest cosmological constraints available in
literature (as we can see from Table \ref{table:res}). Compared to the results we present with a combination
of WMAP 5-year data, SDSS LRG power spectrum shape, and $b_g(L)$ data at three redshifts,
better constraints on the sum of neutrino masses have been published with cosmological analyses that also
make use of clustering measurements of the Lyman-$\alpha$ forest \cite{Fogli:2008ig,Seljak:2006bg}.
There is some possibility that
Ly-$\alpha$ statistics may be more subject to uncertainties in both the measurement and the modeling from the theory side.

The strong limit imposed with Ly-$\alpha$ measurements, combined with other cosmological data, that lead to $\sum m_\nu < 0.17$ eV \cite{Seljak:2006bg},
rules out the evidence for a non-zero neutrino mass claimed in Ref.~\cite{Allen:2003pta}, where the
combination of CMB data with 2 degree Field Galaxy Redshift Survey
\cite{Colless:2003wz} data, X-ray luminosity function observations
(XLF) and baryonic gas mass fraction measurements led to the
constrain $\sum m_{\nu}=0.56^{+0.30}_{-0.26}$ eV.
Independent of Ly-$\alpha$ data, with a 95\% confidence level limit on the sum of neutrino masses of 0.28 eV,
we can also conclusively state that this suggested detection of neutrino masses limit is not compatible with a different set of
cosmological data involving galaxy clustering and clustering bias, combined with CMB.
The origin of the result in Ref.~\cite{Allen:2003pta} was due to the inclusion of X-ray measurements that favored low
values of matter amplitude fluctuations, $\sigma_8=0.70\pm0.04$. Due
to the degeneracy between $\sum m_{\nu}$ and $\sigma_8$ from CMB
data  the inclusion of XLF data in the analysis led therefore to the
evidence for a nonzero neutrino mass, that can be ruled out when
considering increased uncertainties in the X-ray data. We also refer the reader to Refs.~
\cite{Allen:2003pta} and \cite{Tegmark:2003ud} for further discussions about
this result.

While our limit on 0.28 eV, in the near future, cosmological data could reach the sensitivity level of $\sim$ 0.1 eV to begin to distinguish
between the normal and the inverted neutrino mass hierarchy. For example, inverted hierarchy could be ruled out if we can exclude that $\sum
m_{\nu}> 2\sqrt(\Delta\,m_{23}^2)$ where $\Delta\,m_{23}^2\sim 2.4 \cdot 10^{-3} eV^2$ is the squared mass difference between neutrino
mass eigenstates \cite{Fogli:2005cq}. We also note that the current limit on the sum of neutrino masses from cosmological observations
could be quite important for ongoing and future experiments which
aim to measure neutrino masses: for example there is a tension
between our results and the limits of $0.16<m_{\beta\beta}<0.52
(2\sigma)$ on the neutrino mass coming from the analysis of part of
the Heidelberg-Moscow experiment on neutrinoless double beta decay
\cite{HM}: moreover, the next-generation tritium $\beta$-decay
experiment KATRIN \cite{KATRIN} would not be able to measure the
absolute value of neutrino mass because its detectability threshold
is at $\sim 0.2$ eV.

\begin{table}[h]
\begin{center}
\begin{tabular}{lrrr|rrr}
\hline
&WMAP5+SDSS\\
&+ACBAR+HST+SNe+all z's\\
&        &     \\
\hline \hline
 $ \Omega_b h^2$     &  $0.02264\pm0.00054$ \\
 $ \Omega_c h^2$     &  $0.1203\pm0.0055$\\
 $ \tau$             &  $0.088\pm0.016$\\
 $ n_s$              &  $0.961\pm0.13$ \\
 $w$                 &  $-1.125\pm0.092$ \\
  $ln(10^{10} A_s)$     &   $3.22\pm0.05$\\
 $ \Omega_m$         &  $0.285\pm0.023$      \\
 $\sigma_8$          &  $0.778\pm0.037$     \\
 $\sum m_{\nu}$      &  $<0.56$eV  (95\%CL)\\
 $ b^*_1$            &  $1.03\pm0.02$   \\
 $ b^*_2$            &  $1.22\pm0.07$  \\
 $ b^*_3$            &  $3.41\pm0.33$  \\
 \hline
\hline
\end{tabular}
\caption{Mean values and $1\sigma$ constraints on cosmological
parameters from WMAP combined with ACBAR experiment, $SNe-Ia$ data, $HST$ prior and all bias data sets.}\label{table:3}\vspace{1cm}
\end{center}
\end{table}

\section{Conclusions}

Here we have used galaxy bias measurements as a function of
luminosity to put constraints on cosmological parameters, mainly
$\sigma_8$, $\sum m_{\nu}$, and the dark energy equation-of-state $w$. We used three galaxy bias-luminosity data sets at three different
redshifts and improved by roughly a factor of two previous constraints obtained with an analogous set of
data, finding the sum of neutrino masses to be $\sum m_{\nu}<0.28$eV at the $95\%$ confidence level for a $\Lambda
CDM +m_{\nu}$ model, with $\sigma_8=0.759\pm0.025$. We also have
shown that redshift evolution of the bias information can constrain the equation-of-state of dark
energy and we obtained the constraints $w=-1.12\pm0.10$ for a
$\Lambda CDM+w$ model and $w=-1.30\pm0.19$ for a $\Lambda CDM
+m_{\nu}+w$ model.

In the case of dark energy equation of state
allowed to vary the constraints on the sum of neutrino masses are weakened to
$\sum m_{\nu}<0.59$eV at the 95\% confidence level, but this still improves previous constraints from WMAP
combined with BAO and SNe  data only ($\sum m_{\nu}<0.66$ eV). While the inclusion of Ly-$alpha$ data have led to stronger constraints
on the sum of neutrino masses, our constraint with a minimal set of data is still competitive and only uses clustering information of galaxies and the primordial fluctuations
as probed by the CMB. In the future methods such as the one we use could further improve constraint on neutrino
masses independent of Ly-$\alpha$ data and achieve the sensitivity necessary to distinguish
between the normal and the inverted neutrino mass hierarchies.

\textit{Acknowledgments}

This work was supported by NSF CAREER AST-0645427. We thank Tristan Smith for useful discussions.
AC thanks Dipartimento di Fisica  and INFN, Universita' di Roma-La Sapienza
and Aspen Center for Physics for hospitality while this research was initiated.
FDB thanks UCI Center for Cosmology for hospitality while this research was conducted.

\end{document}